\documentclass[useAMS,usenatbib,usegraphicx]{mn2e}

\usepackage[english]{babel}
\usepackage[utf8]{inputenc}

\usepackage{times}

\newcommand{\araa}{Ann. Rev. A\&A}
\newcommand{\mnras}{MNRAS}
\newcommand{\aaps}{A\&A}
\newcommand{\aap}{A\&A}
\newcommand{\apjs}{ApJS}
\newcommand{\apj}{ApJ}
\newcommand{\aj}{AJ}
\newcommand{\nat}{Nature}
\newcommand{\apjl}{ApJL}

\title[H$_{2}$O Maser Polarization of the Water Fountains IRAS 15445$-$5449 and IRAS 18043$-$2116]
  {H$_{2}$O Maser Polarization of the Water Fountains IRAS 15445$-$5449 and IRAS 18043$-$2116}
\author[A. F. Pérez-Sánchez et al.]
  {A.~F.~Pérez-Sánchez,$^1$\thanks{email: aperez@astro.uni-bonn.de} 
  W.~H.~T.~Vlemmings,$^1$ J.~M.~Chapman$^2$\\
  $^1$Argelander Institute for Astronomy, University of Bonn,
      Auf dem Hügel 71, 53121 Bonn, Germany\\
  $^2$CSIRO Astronomy and Space Science, Australia Telescope National Facility, PO Box 76,
      Epping, NSW 1710, Australia}

\date{Accepted 2011 August 5.  Received 2011 August 4; in original form 2011 July 20}

\pagerange{\pageref{firstpage}--\pageref{lastpage}} \pubyear{2011}

\begin{document}

\label{firstpage}

\maketitle

\begin{abstract}
We present the morphology and linear polarization of the 22-GHz H$_{2}$O masers
in the high-velocity outflow of two post-AGB sources, d46 (IRAS 15445$-$5449) 
and b292 (IRAS 18043$-$2116). The observations were performed using The Australia
Telescope Compact Array. Different levels of saturated maser emission have been detected
for both sources. We also present the mid-infrared image of d46 overlaid with the 
distribution of the maser features that we have observed in the red-shifted lobe of the bipolar structure. 
The relative position of the observed masers and a previous radio continuum observation suggests that the
continnum is produced along the blue-shifted lobe of the jet. It is likely due to synchrontron radiation, 
implying the presence of a strong magnetic field in the jet. The fractional
polarization levels measured for the maser features of d46 indicate that the polarization vectors are tracing the
poloidal component of the magnetic field in the emitting region.
For the H$_{2}$O masers of b292 we have measured low levels of fractional linear polarization. 
The linear polarization in the H$_{2}$O maser region of this source likely indicates a dominant toroidal or poloidal magnetic
field component. Since circular polarization
was not detected it is not possible to determine the magnetic field strength. 
However, we present a $3$-$\sigma$ evaluation of the upper limit intensity of the magnetic field 
in the maser emitting regions of both observed sources.
\end{abstract}


\section{INTRODUCTION}
Post-Asymtoptic Giant Branch (post-AGB) stars represent a very short phase in the evolution of
low and intermediate initial mass stars ($M_{\star}\la 9M_{\odot}$).
During the post-AGB phase, the high mass-loss rate ($10^{-7}-10^{-4}\rm{M_{\odot} yr^{-1}}$) observed at the 
end of the Asymtoptic Giant Branch (AGB) evolution decreases. Simultaneously, the effective temperature of 
the central star increases, while the circumstellar envelope (CSE) slowly detaches from the star 
(see \citealt{OlofssonBook} for a review). 
The post-AGB phase ends when the central star 
is hot enough to ionize the material which was ejected from the AGB phase, 
forming a new Planetary Nebula (PN) (e.g. \citealt{van}). It is generally assumed  
that the mass-loss process along the evolution
in the AGB is spherically symmetric \citep{OlofssonBook}. However, a high
percentage of PNe have been observed showing aspherical symmetries that include
elliptical, bipolar or multipolar shapes. It is still not clear at
what point in the evolution toward a PNe the departure from the spherical
symmetry starts and even more importantly, what the physical processes
involved to form the complex shapes observed are. Companion interactions, binary sources, magnetic fields, 
fast and slow wind interaction, disks and a combination of these have been considered as 
the main factors to shape PNe (e.g. \citealt{sahai}, \citealt{balick} and references therein).\\
Hydroxyl (OH) and Water (H$_{2}$O) maser emission have been observed in the CSE of the
progenitors of the (pre-)PNe. Typically, for spherically symmetric CSEs, the double peak spectra
of the OH masers defines the most blue- and red-shifted velocities
in the CSE with respect to the stellar velocity. The velocity range of the H$_{2}$O maser spectra is narrower than
the OH velocity distribution, and the H$_{2}$O masers are confined closer to the central star. 
A class of post-AGB stars, the so called ``Water Fountains'', is characterized by the detection of 
H$_{2}$O maser emission over an unusually large velocity range broader than
the velocity range defined by the OH maser emission \citep{likkel}. Sources
with H$_{2}$O maser velocity spread over a range of $\ga100\rm{km\ s^{-1}}$ 
have been detected (e.g. \citealt{likkel}, \citealt{deacontwo}, \citealt{walsh}, \citealt{Gomez}). Those
H$_{2}$O masers have been observed in regions where the interaction between the
high-velocity outflow and the slow AGB wind seems to be active, hence the H$_{2}$O
masers are probably excited in the post-shock region.
Recent infrared imaging of water fountains have revealed bipolar and multipolar morphologies \citep{lagadec}.  
\citet{wouter} have detected circular and linear polarization in the H$_{2}$O
maser features along the jet of W43A, the archetypal water fountain. They found that the jets of W43A are magnetically 
collimated. Therefore, the detection of polarized maser emission from water fountains is useful to determine the role
of the magnetic fields on the onset of wind asymmetries during the evolution from AGB stars to aspherical PNe.\\
Here we report the detection of linear polarization of 22-GHz H$_{2}$O maser emission from two water fountains 
d46 and b292 (IRAS 15445$-$5449, IRAS 18043$-$2116). 

\begin{figure*}
\begin{minipage}{160mm}
  \includegraphics[width=80mm]{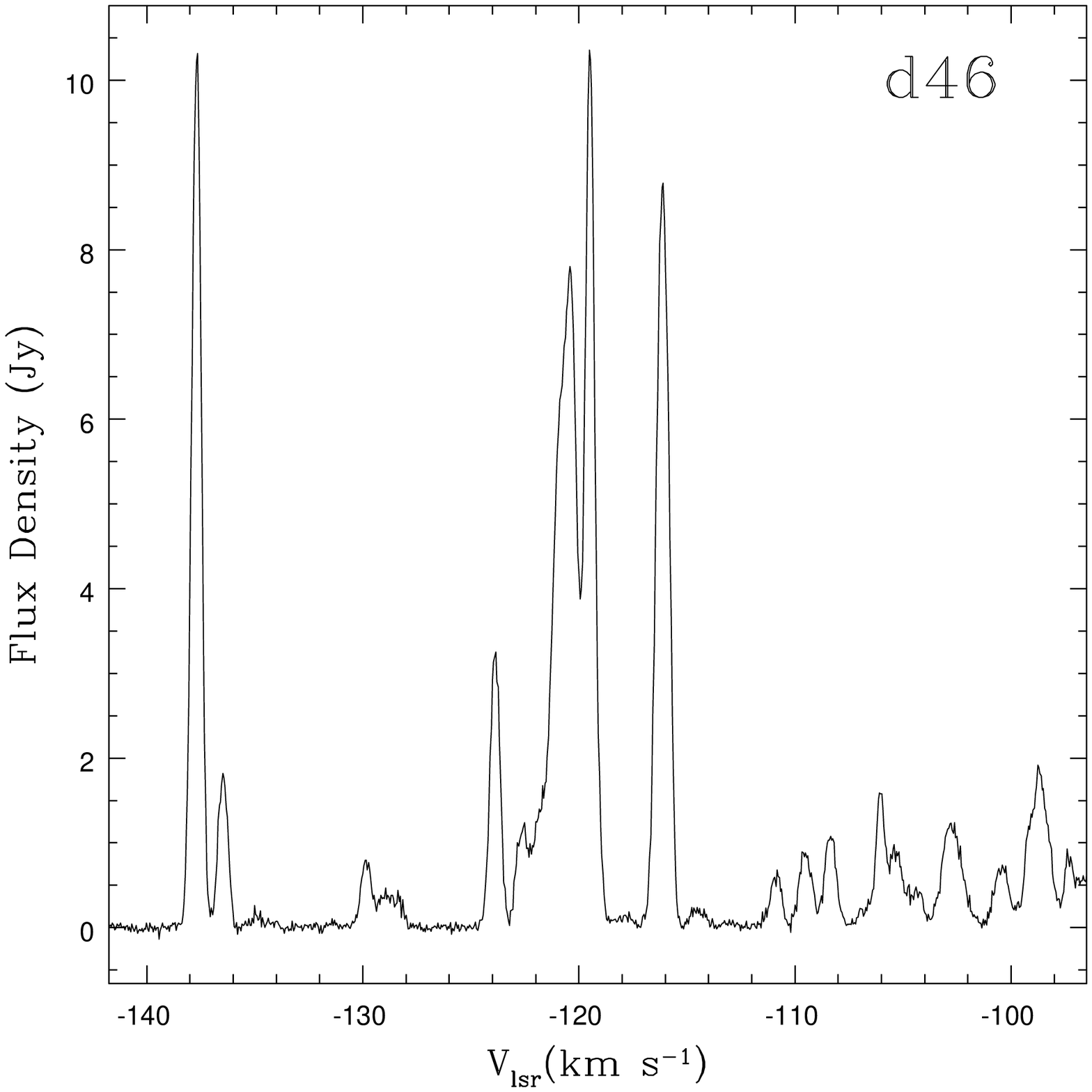}
  \includegraphics[width=80mm]{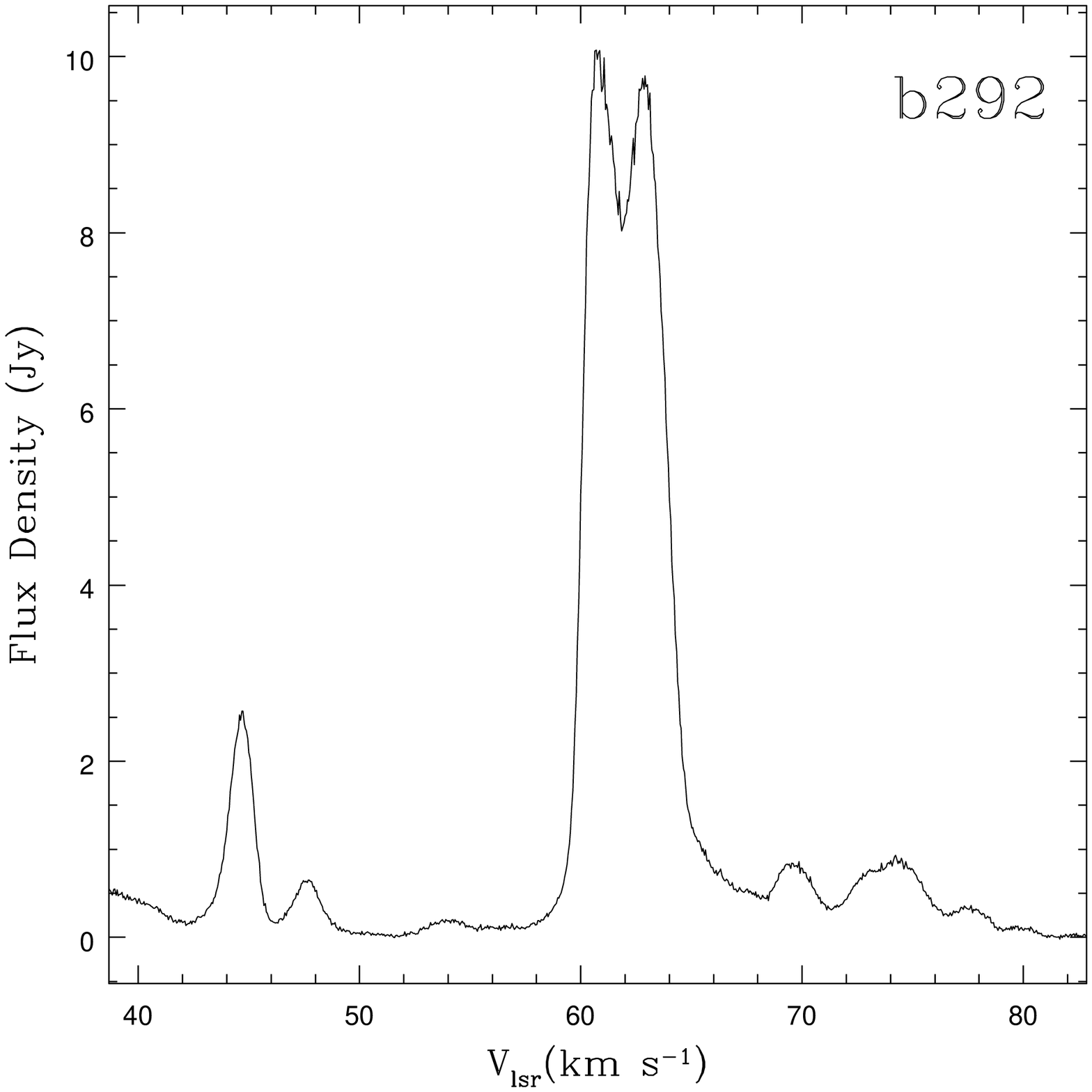}
  \caption{The observed section of the H$_{2}$O maser spectra for: d46 (IRAS 15445$-$5449), and b292 (IRAS 18043$-$2116). 
    The velocity width of the spectra was limited by the bandwidth available at the ATCA at the time of our observations. The 
    observations were centered on the brightest emission peak.}
\label{fig:spectra}
\end{minipage}
\end{figure*}  

\section{OBSERVATION AND DATA REDUCTION}
\label{sec:2} 
The Australia Telescope Compact Array (ATCA) was used to observe the H$_{2}$O maser emission 
from d46 (IRAS 15445$-$5449) on 2006 November 28-30 using the 6B array configuration, and b292 (IRAS 18043$-$2116) 
on 2007 July 6-9 with the 6C array configuration. The full track of observations of d46 was 8 hr, consisting of observations of the
source interspersed with scans on the phase calibrator 1613$-$586. Later, b292 was observed over 16 hr, and 
likewise, its observations were interleaved with a 
phase reference source, 1730$-$130. The primary flux calibrator source, 
1934$-$638, was observed during each observation program. The flux density of 1934$-$638 at 22.2-GHz was taken to be
0.81 Jy. Both observations were performed in full polarization 
mode, with rest frequency of 22.23508-GHz, using 4-MHz bandwidth with 1024 spectral channels, 
covering a velocity width $\sim$50 $\rm{km\ s}^{-1}$ and a velocity resolution of  
$0.05\ \rmn{km\ s}^{-1}$. Before starting the full observations, some
snapshots were carried out, for each source, to determine at what range of velocity the brightest 
emission occurred. As we could not cover the entire velocity range, the observing bands were centered 
on the brightest emission peak.\\
The data reduction was performed using the MIRIAD package \citep{miriad}. After flagging bad
visibilities, the flux density for each phase calibrator was determined using the flux density of 
primary flux calibrator. We obtained flux densities of 1.83 Jy and 2.83 Jy for 1613$-$586 and 
1730$-$130 respectively. Both sources were used to perform the polarization calibration as well as 
to determine amplitude and phase solutions, which were applied to the corresponding target data set. 
For polarization analysis, image cubes for stokes I, Q, U and V were created. 
The resulting noise level in the emission free channels of the I cube of d46
was $\sim$37 $\rm{mJy\ beam^{-1}}$, whereas that of b292 was $\sim$13 $\rm{mJy\ beam^{-1}}$.
The stokes I image cubes were analysed using the AIPS task
SAD, which was used to fit, channel by channel, all the components with peak flux densities 
higher than five times the rms with two-dimensional Gaussians. Only those maser features found in at least five consecutive channels
were considered reliable detections. The Right Ascension and 
Declination for the maser features were calculated as the mean position for each set of 
consecutive channels. Figure \ref{fig:spectra} shows the H$_{2}$O maser spectra for both sources.  

\begin{table*}
 \centering
 \begin{minipage}{140mm}
  \caption{Maser features.}
  \begin{tabular}{ccccccc}
  \hline
  & & & & d46 (IRAS 15445$-$5449) & &\\
  \hline
  \hline
          & Peak Intensity &  RA      &  DEC      &   $V_{lsr}$     &  $P_{L}$ &   $\chi$\\
  Feature &  (Jy $beam^{-1}$) & (15 48 X) & ($-$54 58 X) & (km $s^{-1}$) &  (per cent)   & (deg)\\
  \hline
1   &    11.662 & 19.4063 &	20.165 &	$-$137.6 & 4.1 $\pm$ 0.6 & $-$73.0 $\pm$ 1.6\\
2   &    2.055 & 19.4064 &	20.155 &	$-$136.5 & 7.5 $\pm$ 0.7 & $-$40.7 $\pm$ 4.6\\
3   &	0.909 &	19.4025	&	20.099 &	$-$129.9 &       --	&   --  \\
4   &	0.537 &	19.4086	&	20.090 &	$-$128.9 &       --       &   --  \\
5   &	5.021 &	19.3972 &	19.942 &	$-$123.8 &       --       &   --  \\
6   &	9.077 &	19.4032	&	20.069 &	$-$120.4 & 3.4 $\pm$ 0.9 & $-$51.6  $\pm$ 13.1\\
7\footnote{Reference features in figure \ref{fig:h2omas} and \ref{fig:spowoh}}   & 11.883	&  19.4038 & 20.083 & $-$119.5 & 2.9 $\pm$ 0.5 &	$-$56.5 $\pm$ 4.8\\
8   &	10.334 & 19.4029 &	20.057 &	$-$116.1 & 8.3 $\pm$ 0.9 & 83.4 $\pm$ 2.7\\
9   &	0.351 &	19.3950	 &	19.922 &        $-$114.9 &       --       &     --\\
10  &	0.735 &	19.3736	 &	19.363 &	$-$113.2 &       --	&     --\\
11  &	0.820 &	19.4012	 &	20.041 &        $-$110.8 &       --       &     --\\
12  &	1.166 &	19.3994	 &	19.997 &	$-$109.5 &       --       &     --\\
13  &	1.380 &	19.3897	 &	19.977 &	$-$108.4 &       --       &     --\\
14  &	2.055 &	19.4011	 &	20.011 &	$-$106.4 &       --       &     --\\
15  &	1.619 &	19.3960	 &	19.898 &	$-$105.3 &       --       &     --\\
16  &	2.201 &	19.3564	 &	19.933 &	$-$104.2 &       --       &     --\\
17  &	2.606 &	19.3912	 &	19.851 &	$-$102.8 &       --       &     --\\
18  &	2.922 &	19.3598	 &	19.889 &	$-$100.3 &       --       &     --\\
19  &	7.142 &	19.3593	 &	19.858 &	$-$98.6  &       --       &     --\\
20  &	3.209 &	19.3601	 &	19.824 &	$-$97.3  &       --       &     --\\
\hline
\hline
& & & & b292 (IRAS 18043$-$2116) & &\\
\hline
          & Peak Intensity &  RA      &  DEC      &   $V_{lsr}$     &  $P_{L}$ &   $\chi$\\
  Feature &  (Jy $beam^{-1}$) & (18 07 X) & ($-$21 16 X) & (km $s^{-1}$) &  (per cent)   & (deg)\\
\hline
\hline
1   &   2.820 & 20.8484  &      11.845 &        44.7   &      --	&     --\\
2   &	0.751 &	20.8485  &      11.891 &        47.7   &      --	&     --\\
3   &	0.240 &	20.8500	 &      11.886 &        54.3   &      --	&     --\\
4$^{a}$ &  11.206 & 20.8480 & 11.860  & 60.8  & 1.3 $\pm$ 0.2 & $-$4.2 $\pm$ 4.9\\
5   &	10.821 & 20.8485 &      11.865 &        62.9   & 1.5 $\pm$ 0.2 & $-$6.8 $\pm$ 7.3\\
6   &	0.942  & 20.8483 &      11.844 &        69.4   &      -- 	&     --\\
7   &	0.500  & 20.8488 &      11.835 &        72.0   &      --        &     --\\ 
8   &	1.023  & 20.8477 &      11.838 &        74.3   &      --        &     --\\
9   &	0.667  & 20.8484 &      11.880 &        77.6   &      --        &     --\\
10  &	0.163  & 20.8476 &      12.040 &        79.8   &      --        &     --\\
\hline
\end{tabular}
\label{tab:one}
\end{minipage}
\end{table*}

\section{RESULTS}
\label{sec:3}
\begin{figure*}
\begin{minipage}{160mm}
  \includegraphics[width=80mm]{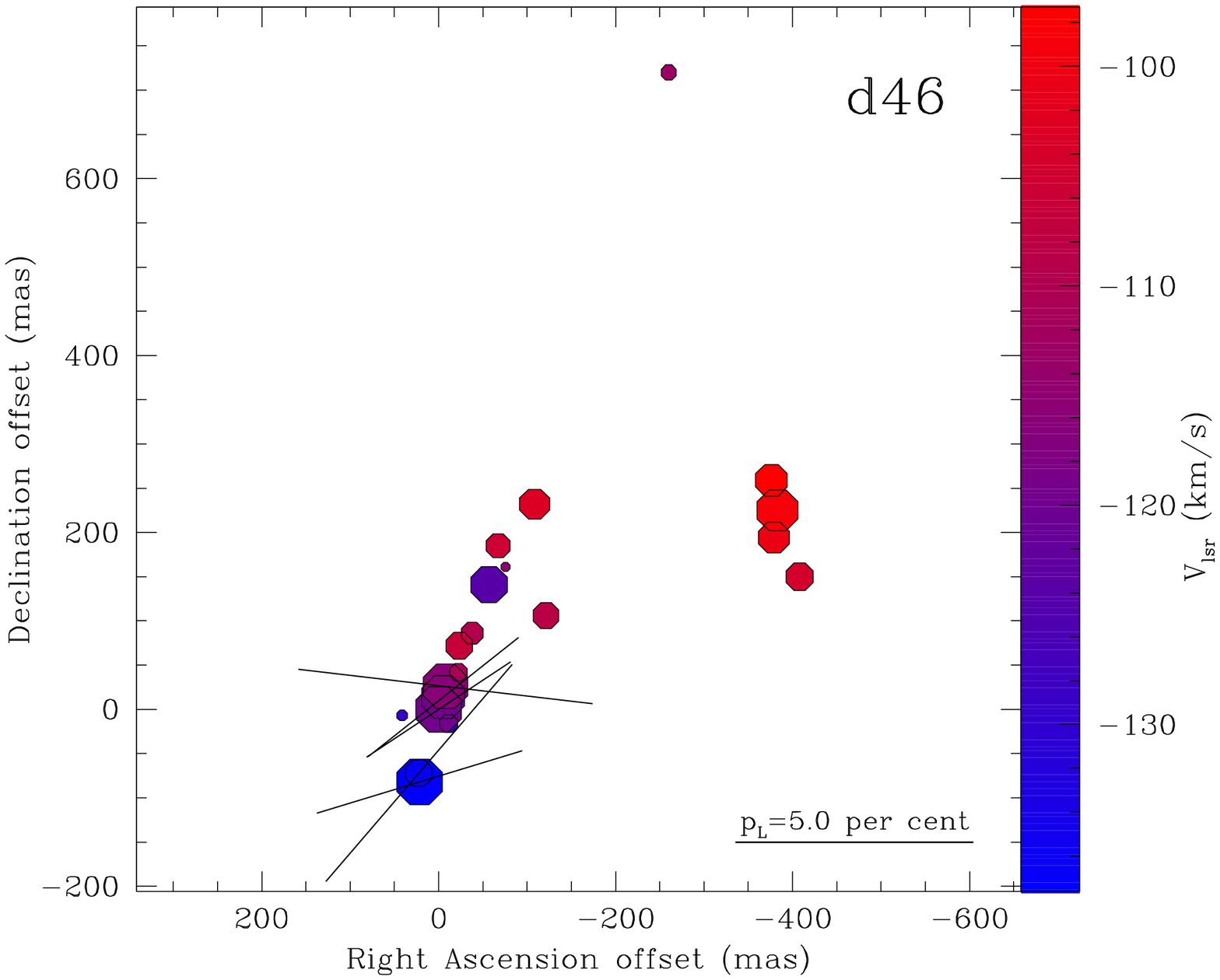}
  \includegraphics[width=80mm]{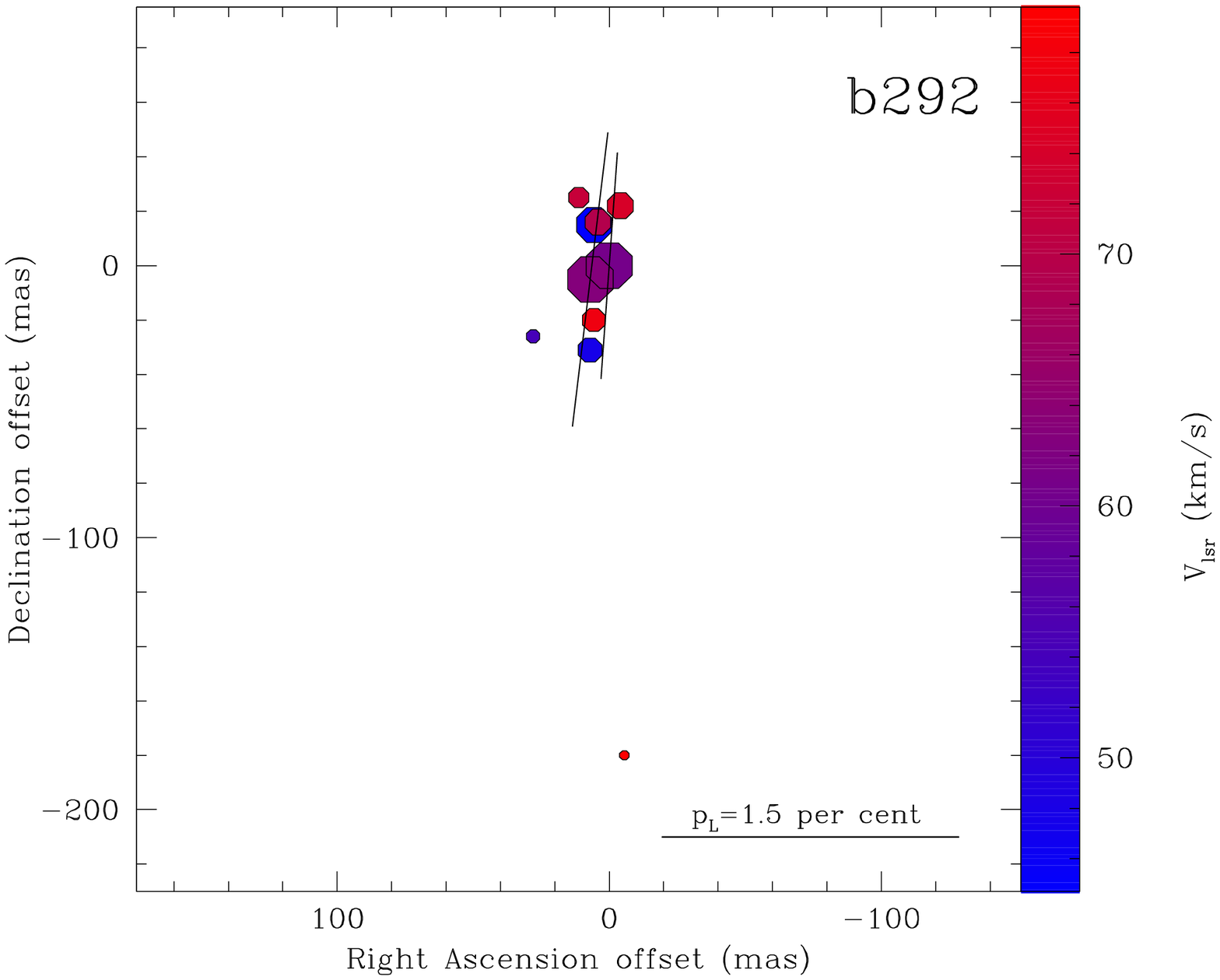}
  \caption{H$_{2}$O maser region of d46 ({\it left}) and b292 ({\it right}). The offset positions are with
  respect to the reference position indicated in table \ref{tab:one} , which are the brightest maser spots in each region. The vectors show
  the polarization angle, whose length is scaled according the linear polarization fraction. The horizontal bar at the bottom of each image
  sets 5 per cent (d46) and 1.5 per cent (b292) of linear polarization fraction.}
\label{fig:h2omas}
\end{minipage}
\end{figure*}

In Table \ref{tab:one}, we present the results of the analysis of the 
H$_{2}$O masers features of d46 and b292. We list the peak
intensity, the position as RA and DEC, LSR velocity ($V_{\rm{lsr}}$), 
fractional linear polarization ($p_{L}$) and polarization angle
($\chi$). In Figure \ref{fig:h2omas} we show 
a map of the H$_{2}$O maser regions for both observed sources.
The size of each feature is scaled according to the peak intensity, the
radial velocity is colour-coded and the vectors show the polarization
angles. We have identified twenty H$_{2}$O maser features for d46, with
emission detected at velocities from $-97.3\ \rm{km\ s}^{-1}$ to $-137.6\ \rm{km\ s}^{-1}$, red-shifted
respect to the systemic velocity $\sim -150\ \rm{km\ s}^{-1}$ \citep{deacontwo}.
Fractional linear polarization has been detected for five of the twenty features, particularly for the most 
blue-shifted features, with levels from $2.9\pm0.5$ per cent to $8.3\pm0.9$ per cent. No significant 
circular polarization was detected.\\ 
Ten H$_{2}$O maser features have been identified in our observation of b292, with emission at velocities 
from $44.7\ \rm{km\ s}^{-1}$ to $79.8\ \rm{km\ s}^{-1}$. \citet{deacone} have derived
the systemic velocity of the source from the double-peaked 1665-MHz spectrum, obtaining
a stellar velocity of $87.0\ \rm{km\ s^{-1}}$. Thus, because of the limited bandwidth,
our observation was restricted to blue-shifted velocities only.  
Fractional linear polarization was detected only for the two brightest features. 
As presented in table \ref{tab:one}, not only are the fractional linear polarization levels for both features
almost the same, but so is the polarization angle. As was the case for d46, 
no significant circular polarization was detected.
\vfill 

\begin{figure*}
\begin{minipage}{160mm}
  \includegraphics[width=80mm]{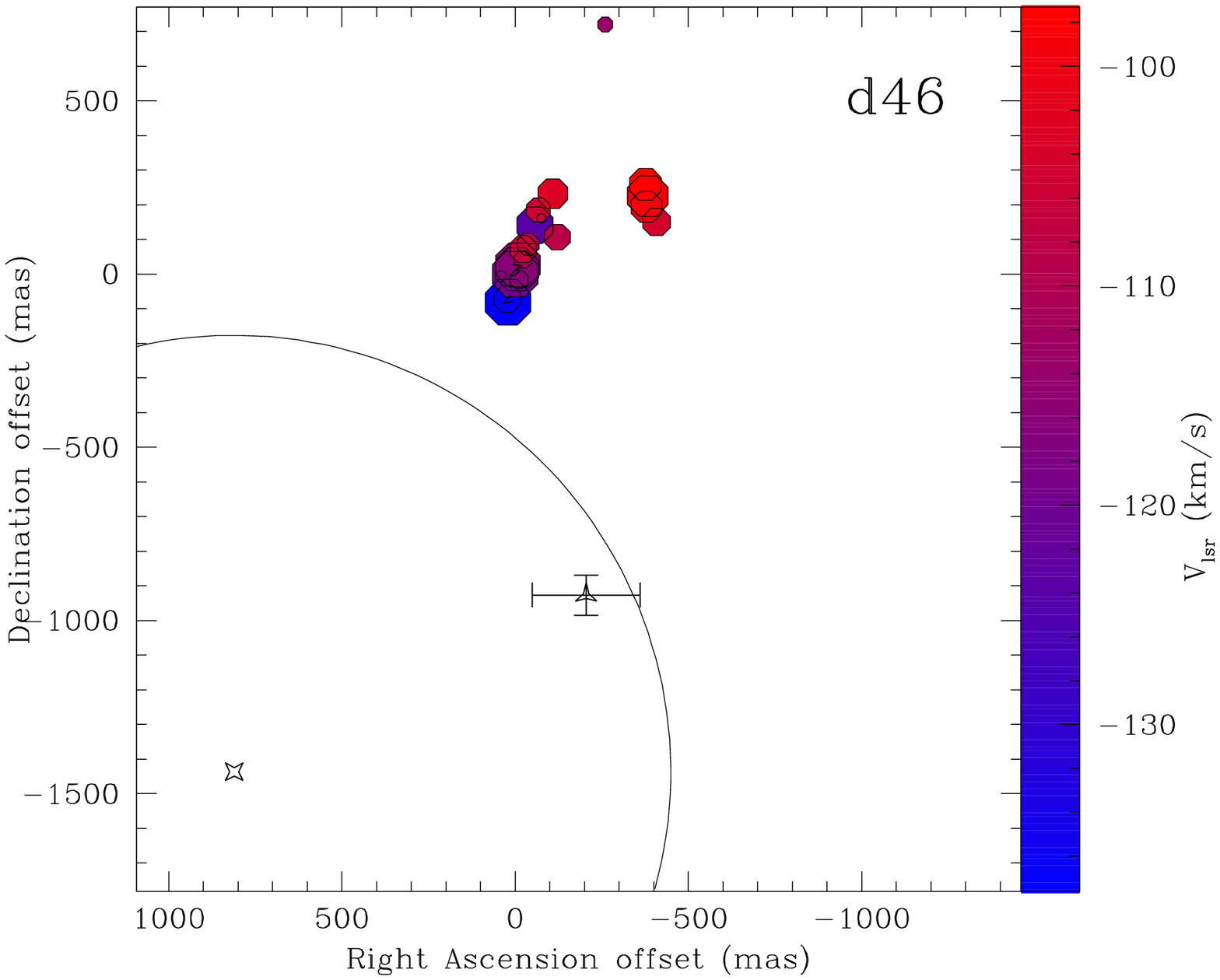}
  \includegraphics[width=80mm]{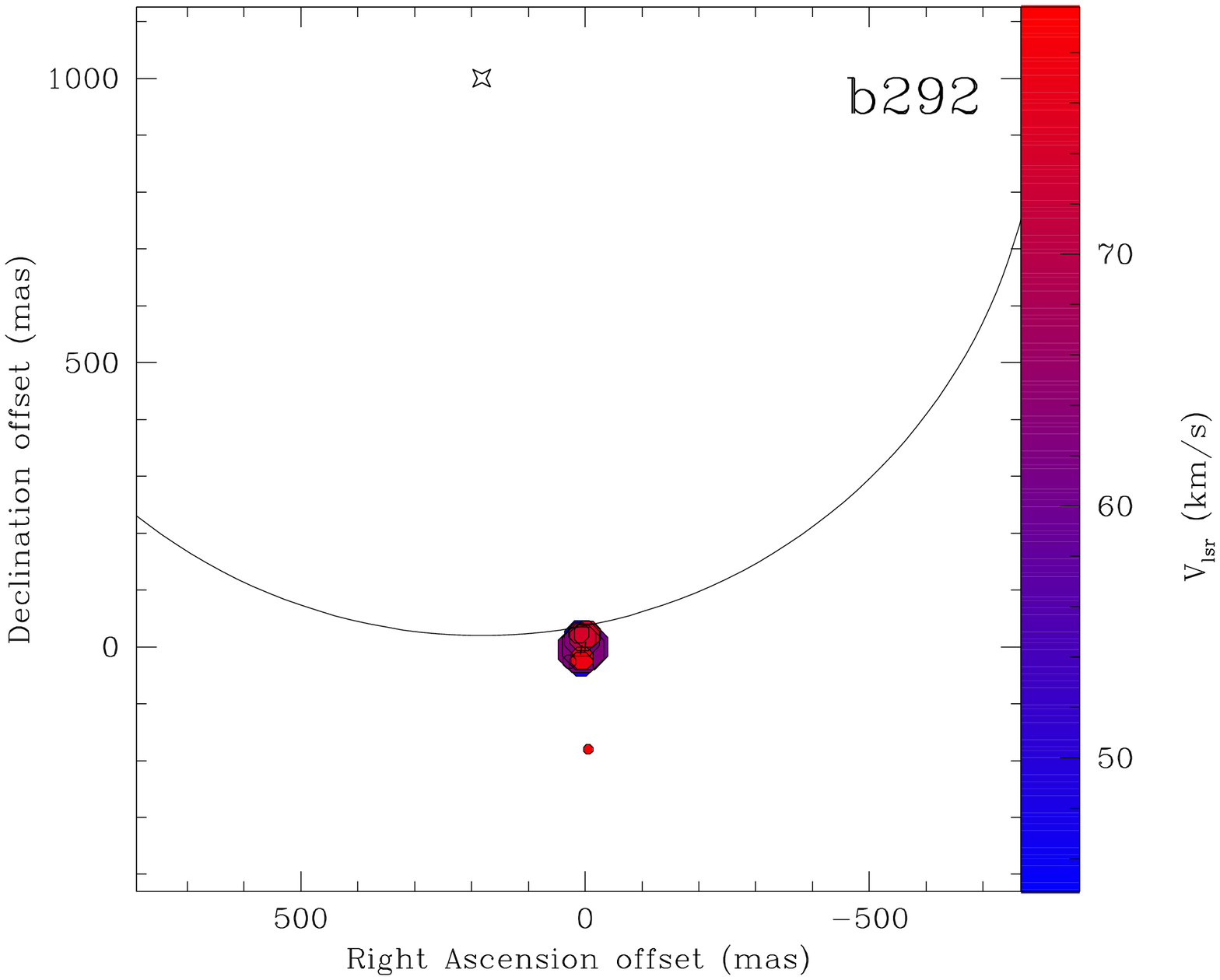}
  \caption{The image in the left shows the position of the radio continuum emission (triangle with error bars) \citep{bains},
    the OH position (star) and its uncertainty (arc) \citep{deacone} with respect to the H$_{2}$O maser features that we have detected for d46. 
    In the same way, the right image shows the OH position (star with four peaks) and its uncertainty (arc) relative to the H$_{2}$O maser features
    from our analysis for b292}
\label{fig:spowoh}
\end{minipage}
\end{figure*}  

\section{DISCUSSION}
\label{sec:4}
\subsection{d46 ({\it IRAS 15445$-$5449, OH326.530$-$00.419})}
\label{subsec:4.1}
\subsubsection{Previous Observations}
According to the MSX ([8$-$12],[15$-$21]) and IRAS ([12$-$25],[25$-$60]) two-colour diagrams, d46 has been classified as a 
highly-evolve late post-AGB object (\citealt{sevemsx}, \citealt{deacone}, \citealt{bains}). After 22-GHz H$_{2}$O maser 
emission was detected with a velocity range 
beyond that of the OH masers, it was classified as a water fountain source \citep{deacontwo}.\\
Both main-line and 1612-MHz OH maser emission have been detected toward
d46 (\citealt{catagtwo}, \citealt{deacone}). Based on the irregular
line profile of the three lines detected, \citet{deacone} suggested
that the source was likely a bipolar object, since irregular OH maser profiles
are associated with sources that almost certainly have bipolar outflows and a remnant
AGB wind. This hypothesis was confirmed by the Mid-infrared images (11.85 and 12.81 $\rm{\mu m}$)
of the source, obtained with VISIR/VLT and recently published by \citet{lagadec}. 
From the images the bipolar structure of the object is clear, with
dust emission around the jet lobes. Indeed, the infrared photometry and the IRAS LRS 
data reported in the literature (\citealt{bains}, \citealt{lagadec}) show that the contribution
at short wavelengths, which is due to the photospheric component,
is still weak and the double-peaked infrared SED, characteristic
of post-AGB objects, is mainly dominated by the dust emission. Therefore,
the central star is surrounded by an optically thick structure \citep{lagadec}
which could be a dense equatorial torus.\\
Non-thermal radio continuum emission has been detected at 3, 6 and 13 cm
(\citealt{cohen}, \citealt{bains}) with significant flux variation among the observations
(spectral index $\alpha \approx -0.8$ \citealt{cohen}, $\alpha \approx -0.34$ \citealt{bains}).
Such detections have been associated with synchrotron emission produced by shocks
between the high-velocity wind and the slow AGB remnant. The flux variation has
been interpreted as episodic shocks between both wind components. 

\subsubsection{The H$_{2}$O masers}
We have observed H$_{2}$O maser features
only from the red-shifted velocities with respect to the OH maser features, 
which are centered at $\sim -150\ \rm{km\ s^ {-1}}$
\citep{deacontwo}. The brightest features were detected between $-$115.0 and $-$140.0 km $\ \rm{s^ {-1}}$, 
considering the overlapped region, the overall H$_{2}$O spectrum that we have detected 
is quite similar to the spectrum detected by \citet{deacontwo}, although as expected with some difference 
of the flux densities. Figure \ref{fig:h2omas} ({\it left}) shows that the projected spatial distribution of the H$_{2}$O
maser features resembles a bow-shock-like structure, similar to that observed by \citet{boboltz} for the
water fountain OH 12.8$-$0.9. Figure \ref{fig:spowoh} ({\it left}) shows both the OH and radio continuum
positions with respect to the projected distribution of the 22-GHz H$_{2}$O maser features that we have detected.
The solid circle represents the uncertainty on the OH position, the error bars show
the accuracy of the radio continuum position, whereas the error on the H$_{2}$O maser position of each feature
is within the size of the symbols. The offset of the radio continuum position with respect to that of the 
H$_{2}$O maser agrees very well with the mid-infrared image, and it seems likely that both 
emitting region are at different points along the jet. In fact, since the 
H$_{2}$O maser features have been produced at the red-shifted side of the jet, the position
of the radio continuum could be related to the position of the central star. But, because
the spectral index of the radio continuum implies non-thermal emission \citep{deacontwo} 
it is more likely synchrotron radiation produced at a region along the jet. This suggests 
the presence of a strong magnetic field along the axis defined by the direction of the high-velocity outflow.
In Figure \ref{fig:superima} we show the Mid-Infrared image of d46 recently published by \citet{lagadec}.
The corrected position of the center of the mid-infrared image is RA 15 48 19.420, DEC $-$54 58 20.100, 
with an error circle of radius 2 arcsec (Lagadec, private comunication).   
Thus, as the relative positions are not known accurately enough, for illustration we have overlaid the H$_{2}$O maser 
features with the red-shifted lobe. The offset of the continuum emission suggest 
that it could arise from the blue-shifted lobe, as an effect of a strong magnetic field, 
as mentioned above. Although the error of the OH position encloses part of the blue-shifted lobe, 
it is likely produced in the outer shells of the CSE where the gas has been accelerated to high velocities through
wind-wind collisions.

\subsubsection{Magnetic field}
Two of the maser features were detected to have a high percentage of linear
polarization ($p_{L}>5$ per cent), which likely is a result of maser emission in the 
saturated regime \citep{profe}. For H$_{2}$O masers, the percentage of linear polarization depends on the
degree of saturation of the emiting region and on the angle ($\theta$) between
the direction of the maser propagation and the magnetic field lines. Additionally, there is
a critical value $\theta_{\rm{crit}}$, such that if $\theta\la \theta_{\rm{crit}}= 55^{\circ}$, the polarization 
vectors are parallel to the magnetic field lines, otherwise they both are perpendicular.
According to the $p_{L}$ levels measured, the polarization vectors in figure \ref{fig:h2omas} 
should be perpedicular to the magnetic field lines. The vectors then appear to trace
the poloidal field component, ie, along the direction of the high-velocity
outflow. We did not detect circularly polarized emission. As the masers are relatively weak the 
resulting 3-$\sigma$ magnetic field limit is high, 
$|B_{\parallel}|< 470$ mG.

\subsection{b292 ({\it IRAS 18043$-$2116, OH009.1$-$0.4})}
\label{subsec:4.2}
\subsubsection{Previous Observations}
According to its position on the ([8]$-$[12],[15]$-$[21]) MSX two-colour diagram \citep{sevemsx}, 
b292 is likely to be a young post-AGB object (\citealt{sevemsx}, \citealt{deacone}) . 
This source was confirmed as a water fountain by \citet{deacontwo}, who
have detected 22-GHz H$_{2}$O maser emission over a wide velocity range
of $\approx 210\ \rm{km\ s^{-1}}$. In later observations performed by \citet{walsh}
H$_{2}$O maser emission over the even larger velocity range of $398\ \rm{km s^{-1}}$ has been detected,
one of the largest velocity spread in any Galactic H$_{2}$O maser source. Until the recent observation
of \citet{Gomez}, it was the water fountain with the largest range in H$_{2}$O maser velocities. In addition, 
b292 was the first post-AGB object discovered with emission from the 1720-MHz OH 
satellite line \citep{seven01}. The 1720-MHz OH transition is produced in regions with  
special conditions of temperature and density and it is often associated 
with both Galactic star-forming regions (SFRs) and Supernova remnants \citep{lockett}. 
Its detection is interpreted as an indicator of the existence of C-shocks in the emitting regions. 
The detection of OH maser emission at 1665-MHz and 1612-MHz associated
with the position of the 1720-MHz maser region did confirm that all the OH maser
transitions are related to the same stellar source \citep{deacone}. In fact, the position offset from
the 1665-MHz and 1612-MHz emitting region is less than 0.3'' \citep{seven01}. Neither \citet{seven01} nor  
\citet{deacone} have detected the OH 1667-MHz maser transition, which is unusual for a source
having maser emission at 1665-MHz.

\subsubsection{The H$_{2}$O masers}
As mentioned above, \citet{walsh} have detected H$_{2}$O maser emission at both blue-shifted
and red-shifted velocities relative to the systemic velocity. The projected spatial distribution
and position of such features within the same velocity range  (e.g. Figure 2 of \citet{walsh}) 
is quite consistent with the projected position of those features that we have detected (Figure \ref{fig:h2omas}).
But, considering that the observations were carried out with a time period of almost a year in between, 
it is important to point out the strong variability of the maser flux intesity. 
In our spectra (figure \ref{fig:spectra} {\it right}) there is no maser emission at $53.1\ \rm{km s^{-1}}$, 
while \citet{walsh} detected the brightest peak at that velocity. Besides, they have detected our brightest 
feature at only a half the intensity.\\
As pointed out by \citet{walsh}, the spread in the projected distribution of the maser features with the most extreme 
velocities could be caused by a very small angle between the jet propagation 
direction and the line-of-sight. In addition, \citet{seven01} have pointed out that the lack of a second (red-shifted) 
peak of the 1720-MHz OH maser emission might be because the line-of-sight 
is almost parallel to the jet propagation direction. Although high-resolution images have not
been obtained, we can thus assume that we are looking at the emerging
jet almost pole-on. From \citet{walsh}, the projected orientation in the sky of the jets is East-West. 
The 22-GHz H$_{2}$O maser transition is probably excited
by the shock front of the high-velocity outflow, which hits the slow expanding
AGB envelope. \citet{Surcis} have reported detection 
of 22-GHz H$_{2}$O maser emission produced in regions under C-shock conditions in a high mass SFR.
In late-type stars, C-shocks should be produced at the tip of the emergent outflows,
whereas the post-shock region seems to achieve the necessary conditions to produce the 
1720-MHz OH maser emission. Figure \ref{fig:spowoh} ({\it right}) indicates the position of the 
OH maser reported by \citet{catagone} relative to the 22-GHz H$_{2}$O maser features identified in our analisys. 
The uncertainty of the OH maser position is represented by the solid circle. 
\subsubsection{Magnetic field}
The fractional polarization level detected for the brightest features is low ($<5$ per cent) 
and corresponds to non-saturated H$_{2}$O maser emission. Consequently, the polarization 
vectors could be either perpendicular or parallel to the magnetic field component projected
in the sky plane. Considering the projected jet direction is East-West, this implies that the B-field is either 
almost exactly parallel or perpendicular to the jet. The field in the H$_{2}$O maser region of b292 is thus potentially 
toroidal, as observed for W43A \citep{profe} or poloidal as seen in d46. Significant levels of circular polarization
were not detected. The 3-$\sigma$ upper limit for the field strength is $|B_{\parallel}|< 175$ mG.

\begin{figure}
\begin{minipage}{80mm}
  \includegraphics[width=80mm]{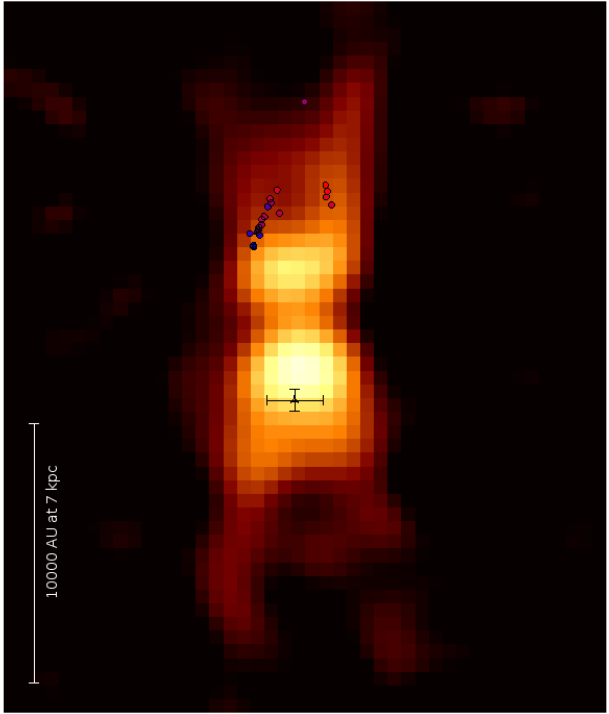}
  \caption{H$_{2}$O maser and radio continuum position overlaid on the mid-infrared image published
    by \citet{lagadec}. For illustration, we have shifted the mid-infrared image, within the 2 arcsec
    of uncertainty of its position, in order of align the H$_{2}$O maser features with the red-shifted 
    side of the high-velocity outflow.}
\label{fig:superima}
\end{minipage}
\end{figure}  

\section{CONCLUSIONS}
\label{sec:5}
We have used the ATCA to observe high-velocity H$_{2}$O emission from two
late-type stars, d46 and b292. We have presented new high angular resolution images of the
H$_{2}$O maser emission in parts of the jet and measured linear polarization
for both sources. The first H$_{2}$O maser maps of d46 show a bow-shock morphology similar to that of 
OH 12.8$-$0.9. According to the level of $p_{L}$ that we have measured, 
the maser emission is in the saturated regime, and the polarization vectors should be
perpendicular to the magnetic field lines, which consequently are the 
poloidal component. In Figure \ref{fig:superima} we have overlaid the position of the masers
with the red-shifted lobe of the jet on the mid-infrared VLTI image of d46 from \citet{lagadec}. The relative position of the
radio continuum emission with respect to those maser features suggest that it is arising from
the blue-shifted lobe and is likely due to synchrotron radiation, indicating the potential presence of
a significant magnetic field in the jet.
Circular polarization was not detected, and we cannot infer the magnetic field
strength along the line-of-sight to better than a 3-$\sigma$ limit of
$|B_{\parallel}|< 470$ mG for d46. New polarization observations, including
that of the radio continuum emission and a broader velocity range 
of the H$_{2}$O spectrum could give accurate measurements
of the magnetic field strength along the jets of d46.
For the H$_{2}$O of b292 we have measured low $p_{L}$ levels,
and the polarization vectors could be either parallel or perpendicular
to the magnetic field lines projected on the sky plane. In fact, the projection
of the jets in the sky is likely East-West, and the polarization vectors
could be associated to either the poloidal (i.e. East-West) or toroidal (i.e. North-South) component
of the B-field. Also in this case, no significant levels of circular polarization were not detected.
The 3-$\sigma$ upper limit for the magnetic field strength,
is $|B_{\parallel}|< 175$ mG.

\section*{Acknowledgments}
The Australia Telescope Compact Array is part of the Australia Telescope National Facility which is 
funded by the Commonwealth of Australia for operation as a National Facility managed by CSIRO.
This research was supported by the Deutsche Forschungsgemeinschaft (DFG) through the Emmy Noether
Research grant VL 61/3-1.

\label{lastpage}

\end{document}